\documentclass[runningheads]{svmult}
\usepackage{makeidx}
\usepackage{graphicx}

\usepackage{subeqnar}
\usepackage{multicol}
\usepackage{physmubb}
\usepackage{amssymb}
\usepackage{amsmath}
\makeindex
\begin{document}
\title*{Linearized Boltzmann Equation and Hydrodynamics for Granular Gases}
\titlerunning{Linearized Boltzmann Equation}

\author{J. Javier Brey \inst{1}
\and James W. Dufty\inst{2} \and Mar\'{\i}a J.
Ruiz-Montero\inst{1}}
\authorrunning{J. Javier Brey et al.}
\institute{F\'{\i}sica Te\'{o}rica, Facultad de F\'{\i}sica,
Universidad de Sevilla, E-41080, Sevilla, Spain.
e-mail:brey@us.es, majose@us.es \and Department of Physics,
University of Florida, Gainesville, Florida 32611, USA. e-mail:
dufty@phys.efl.edu}
\maketitle
\begin{abstract}
The linearized Boltzmann equation is considered to describe small
spatial perturbations of the homogeneous cooling state. The
corresponding macroscopic balance equations for the density,
temperature, and flow velocity are derived from it as the basis
for a hydrodynamic description. Hydrodynamics is defined in terms
of the spectrum of the generator for the dynamics of the
linearized Boltzmann equation. The hydrodynamic eigenfunctions and
eigenvalues are calculated in the long wavelength limit. The
results allow identification of the hydrodynamic part of the
solution to the linearized Boltzmann equation. This contribution
is used to calculate the fluxes in the macroscopic balance
equations, leading to the Navier-Stokes equations and associated
transport coefficients. The results agree with those obtained
earlier by the Chapman-Enskog method. The implications of this
analysis for application of methods of linear response to granular
fluids and derivation of Green-Kubo expressions for transport
coefficients are discussed.

\end{abstract}
\section{Introduction}
The derivation of hydrodynamic equations for granular gases and
identification of expressions for the associated transport
coefficients has been a problem of theoretical and practical
interest for the last two decades
\cite{BDJyR85,BDLSJyCh84,BDGyS95,BDSyG98,BDBDKyS98,BDGyD99,BDGyD02}.
Early direct extensions of the Chapman-Enskog method to inelastic
systems have been refined recently so that the transport
coefficients for the Boltzmann gas are now known accurately
\cite{BDBDKyS98,BDByC01}. These refinements have clarified the
role of the time dependent reference state, the local homogeneous
cooling state (HCS), and its effects on the qualitative and
quantitative characterization of the conditions for the validity
of a hydrodynamic description. The extension of these results to
higher densities also has been accomplished in the context of the
phenomenological Boltzmann-Enskog equation \cite{BDGyD99}.
However, more general conclusions regarding transport in dense
granular fluids are quite limited. On the other hand, for elastic
collisions the analysis of transport in fluids is quite
sophisticated using, for instance, methods of linear response
theory \cite{BDMc89}. Such methods provide formally exact
expressions for transport properties in terms of time dependent
fluctuations in the equilibrium reference state. These
representations have proven to be particularly useful due to the
possibility of their direct evaluation by molecular dynamics
simulation. In this context, it is natural to ask to what extent
the methods of linear response theory can be applied to granular
fluids.

In applications of linear response for fluids with elastic
collisions, the local conserved densities of mass, energy, and
momentum play a special role. They represent the slow variables as
the system approaches uniformity. More significantly, they are the
hydrodynamic modes in the long wavelength limit, i.e the formal
eigenfunctions of the Liouville operator corresponding to the
hydrodynamic eigenvalues. Standard linear response methods
consider perturbations of the homogeneous state (the Gibbs
ensemble) that are linear combinations of the conserved densities.
In this way, hydrodynamic excitations are selected and, by formal
manipulation of the response functions, Green-Kubo expressions for
the transport coefficients are derived directly. A quite
reasonable extension of this method to granular systems, using
linear combinations of the local conserved densities as initial
perturbations, has been described in Ref.\cite{BDGyvN00}. The
resulting Green-Kubo expressions are similar to those for systems
with elastic collisions, but they do not agree at low density with
those from the Chapman-Enskog method noted above. This difference
has been made more precise by reformulating the Chapman-Enskog
results in a Green-Kubo representation \cite{BDDyB02}. The
resulting time correlation functions are not simply those for the
fluxes of the conserved densities, as it is the case for elastic
collisions. The reason is that the hydrodynamic modes for granular
systems are no longer simply related to the conserved densities,
even in the long wavelength limit. Instead, the hydrodynamic modes
and conserved densities are conjugate sets in the sense that the
former generate the hydrodynamic excitation while the latter
measure the response. An objective here is to clarify this subtle
difference in the context of the linearized Boltzmann equation for
inelastic hard particles, and to discuss the consequences for more
general applications.

Clearly, an important first issue is the unambiguous definition of
hydrodynamic modes. This is accomplished by considering the
eigenvalue problem for the operator generating the dynamics for
the Boltzmann equation, linearized around the reference
homogeneous cooling state (see Section \ref{BDs2} below). The
hydrodynamic part of its spectrum is identified first in the long
wavelength limit. Hydrodynamics refers to systematic
approximations to the macroscopic balance equations for the local
relevant densities (here our choice is the deviations of number
density, temperature, and flow velocity from their values in the
homogeneous state). Consequently, we first derive these equations
and find they are closed without approximation in the long
wavelength limit. These asymptotic equations determine $d+2$
hydrodynamic eigenvalues, where $d$ is the dimensionality of the
system. For normal fluids these are all zero, while for granular
systems there are three distinct eigenvalues, one of which is
$d$-fold degenerate. The corresponding eigenvalues for the
linearized Boltzmann equation then identify its asymptotic
hydrodynamic spectrum. The hydrodynamic spectrum more generally is
defined to be the eigenvalues that are continuously connected to
these asymptotic values in the long wavelength limit. The
expression ``hydrodynamic mode'' means the hydrodynamic
eigenfunction with an exponential time dependence, in the
appropriate time units, characterized by the corresponding
hydrodynamic eigenvalue. A primary new contribution here is the
calculation of these hydrodynamic eigenfunctions in the long
wavelength limit, showing their difference from the local
conserved densities.

With the hydrodynamic eigenfunctions identified, the hydrodynamic
part of the solution to the linearized Boltzmann equation can be
extracted. This is the dominant contribution for large times and
long wavelengths. More specifically, this hydrodynamic part is
calculated to first order in the gradients of the hydrodynamic
fields and the result is used to calculate the average fluxes in
the macroscopic balance equations at finite wavelengths. This
provides the Navier-Stokes hydrodynamics for granular gases, and
allows identification of the transport coefficients. The results
agree with those from the Chapman-Enskog method
\cite{BDBDKyS98,BDByC01} and the Green-Kubo representation of Ref.
\cite {BDDyB02}. However, their derivation here exposes more
clearly the hydrodynamic eigenfunctions and their role as the
origin of key differences from gases with elastic collisions.

In the next Section, the nonlinear Boltzmann equation for a
granular gas is defined, and the exact balance equations for the
density, temperature, and flow velocity are derived from it. The
homogeneous state of these equations corresponds to global
cooling, and the corresponding homogeneous cooling solution (HCS)
to the nonlinear Boltzmann equation is described. This solution
represents the homogeneous state about which small spatial
perturbations are considered. The corresponding linear Boltzmann
equation and linear balance equations for these small
perturbations are then described. Next, the eigenvalue problem for
the generator of the time dependence in the linearized Boltzmann
equation is defined in Section \ref{BDs3}, and the hydrodynamic
eigenfunctions and eigenvalues are calculated in the long
wavelength limit. These results are used in Section \ref{BDs4} to
derive the Navier-Stokes hydrodynamic equations, with expressions
for the transport coefficients in terms of the linearized
Boltzmann collision operator. Their equivalence with previous
results from the Chapman-Enskog method is demonstrated, and these
expressions are also given a Green-Kubo representation. Finally,
the analysis is recapitulated, and implications for a more general
treatment of transport in rapid granular fluids beyond the low
density context are discussed.

\section{Nonlinear Boltzmann Equation and the Homogeneous Cooling
State} \label{BDs1} The Boltzmann equation for a system of
inelastic smooth hard disks ($d=2$) or spheres ($d=3$) of mass $m$
and diameter $\sigma $ has the form \cite{BDGyS95,BDBDyS97}
\begin{equation}
\left( \frac{\partial }{\partial t}+\mathbf{v}\cdot \nabla \right)
f(\mathbf{ r},\mathbf{r},t)=J[f|f],  \label{BD2.1}
\end{equation}
where $J[f|f]$ is the nonlinear inelastic Boltzmann collision
operator,
\begin{equation}
J[f|f]=\int
d\mathbf{v}\,\bar{T}(\mathbf{v},\mathbf{v}_{1})f(\mathbf{r},
\mathbf{v},t)f(\mathbf{r},\mathbf{v}_{1},t).  \label{BD2.2}
\end{equation}
Here $\bar{T}(\mathbf{v},\mathbf{v}_{1})$ is the inelastic binary
collision operator
\begin{equation}
\bar{T}(\mathbf{v},\mathbf{v}_{1})=\sigma ^{d-1}\int d\hat{
\boldsymbol{\sigma}}\,\Theta \left( \mathbf{g}\cdot
\hat{\boldsymbol{\sigma}} \right) \mathbf{g}\cdot
\hat{\boldsymbol{\sigma}}\left( \alpha ^{-2}b_{
\boldsymbol{\sigma}}^{-1}-1\right) ,  \label{BD2.3}
\end{equation}
with $\mathbf{g}=\mathbf{v}-\mathbf{v}_{1}$, $\Theta $ the
Heaviside step function, $\hat{\boldsymbol{\sigma}}$ a unit vector
joining the center of the two particles at contact, $\alpha $ the
coefficient of normal restitution, and
$b_{\boldsymbol{\sigma}}^{-1}$ an operator replacing all the
velocities $\mathbf{v}$ and $\mathbf{v}_{1}$ appearing to its
right by their precollisional values $\mathbf{v}^{\ast }$ and
$\mathbf{v}_{1}^{\ast }$ , given by
\begin{equation}
b_{\boldsymbol{\sigma}}^{-1}\mathbf{v}=\mathbf{v}^{\ast
}=\mathbf{v}-\frac{ 1+\alpha }{2\alpha }\left(
\hat{\boldsymbol{\sigma}}\cdot \mathbf{g}\right)
\hat{\boldsymbol{\sigma}},\quad
b_{\boldsymbol{\sigma}}^{-1}\mathbf{v}_{1}= \mathbf{v}_{1}^{\ast
}=\mathbf{v}_{1}+\frac{1+\alpha }{2\alpha }\left( \hat{
\boldsymbol{\sigma}}\cdot \mathbf{g}\right)
\hat{\boldsymbol{\sigma}}. \label{BD2.4}
\end{equation}
For arbitrary functions $g(\mathbf{v},\mathbf{v}_{1})$ and
$h(\mathbf{v}, \mathbf{v}_{1})$, it is
\begin{equation}
\int d\mathbf{v}\int d\mathbf{v}_{1}
g(\mathbf{v},\mathbf{v}_{1})\bar{T}(
\mathbf{v},\mathbf{v}_{1})h(\mathbf{v},\mathbf{v}_{1})=\int
d\mathbf{v} \int d\mathbf{v}_{1}
h(\mathbf{v},\mathbf{v}_{1})T(\mathbf{v},\mathbf{v}
_{1})g(\mathbf{v},\mathbf{v}_{1}),  \label{BD2.5}
\end{equation}
where
\begin{equation}
T(\mathbf{v},\mathbf{v}_{1})=\sigma ^{d-1}\int
d\hat{\boldsymbol{\sigma}} \,\Theta \left( \mathbf{g}\cdot
\hat{\boldsymbol{\sigma}}\right) \mathbf{g} \cdot
\boldsymbol{\hat{\sigma}}\left( b_{\boldsymbol{\sigma}}-1\right) .
\label{BD2.6}
\end{equation}
The operator $b_{\boldsymbol{\sigma}}$ changes the velocities
$\mathbf{v}$ and $\mathbf{v}_{1}$ to its right into their
postcollisional values $\mathbf{ v}^{\prime }$ and
$\mathbf{v}_{1}^{\prime }$,
\begin{equation}
b_{\boldsymbol{\sigma}}\mathbf{v}=\mathbf{v}^{\prime
}=\mathbf{v}-\frac{ 1+\alpha }{2}\left(
\hat{\boldsymbol{\sigma}}\cdot \mathbf{g}\right) \hat{
\boldsymbol{\sigma}},\quad
b_{\boldsymbol{\sigma}}\mathbf{v}_{1}=\mathbf{v} _{1}^{\prime
}=\mathbf{v}_{1}+\frac{1+\alpha }{2}\left( \hat{
\boldsymbol{\sigma}}\cdot \mathbf{g}\right)
\hat{\boldsymbol{\sigma}}. \label{BD2.7}
\end{equation}
By using Eq.\ (\ref{BD2.5}), the equalities
\begin{equation}
\int d\mathbf{v}\,\left(
\begin{array}{c}
1 \\
m\mathbf{v} \\
\frac{mv^{2}}{2}
\end{array}
\right) J[f|f]=\left(
\begin{array}{c}
0 \\
\mathbf{0} \\
-\frac{pd}{2}\zeta \lbrack f]
\end{array}
\right) .  \label{BD2.8}
\end{equation}
are easily obtained. In the above expressions, $p=nT$ is the
pressure (Boltzmann's constant has been set equal to unity as
usual), $n$ being the number density and $T$ the granular
temperature, and $\zeta $ is a a nonlinear functional of the
distribution function,
\begin{equation}
\zeta \lbrack f]=\frac{(1-\alpha ^{2})\pi ^{\frac{d-1}{2}}m\sigma
^{d-1}}{ 4\Gamma \left( \frac{d+3}{2}\right) pd}\int
d\mathbf{v}\,\int d\mathbf{v}
_{1}\,f(\mathbf{r},\mathbf{v},t)f(\mathbf{r},\mathbf{v}_{1},t)g^{3}.
\label{BD2.9}
\end{equation}

The hydrodynamic fields are defined in the standard way in terms
of velocity moments, i.e.
\begin{equation}
n(\mathbf{r},t)=\int d\mathbf{v}\,f(\mathbf{r},\mathbf{v},t),
\label{BD2.9a}
\end{equation}
\begin{equation}
n(\mathbf{r},t)\mathbf{u}(\mathbf{r},t)=\int
d\mathbf{v}\,\mathbf{v}f( \mathbf{r},\mathbf{v},t), \label{BD2.9b}
\end{equation}
\begin{equation}
\frac{d}{2}n(\mathbf{r},t)T(\mathbf{r},t)=\int
d\mathbf{v}\,\frac{m}{2} V^{2}f(\mathbf{r},\mathbf{v},t),
\label{BD2.9c}
\end{equation}
where $\mathbf{V}=\mathbf{v}-\mathbf{u}$ is the velocity of the
particle relative to the local velocity flow $\mathbf{u}$. Their
balance equations follow directly from Eq.\ (\ref{BD2.8}),
\begin{equation}
\frac{\partial n}{\partial t}+\boldsymbol{\nabla}\cdot \left(
n\mathbf{u} \right) =0,  \label{BD2.10}
\end{equation}
\begin{equation}
\frac{\partial \mathbf{u}}{\partial t}+\mathbf{u}\cdot
\boldsymbol{\nabla} {\bf u}+(mn)^{-1}{\boldsymbol{\nabla}} \cdot
\mathsf{P}=0,  \label{BD2.11}
\end{equation}
\begin{equation}
\frac{\partial T}{\partial t}+\mathbf{u}\cdot
\boldsymbol{\nabla}T+\frac{2}{ nd}\left(
\mathsf{P}:\boldsymbol{\nabla} {\bf u}+\boldsymbol{\nabla}\cdot
\mathbf{q}\right) +T\zeta =0.  \label{BD2.12}
\end{equation}
The functionals giving the pressure tensor $\mathsf{P}$ and the
heat flux $ \mathbf{q}$ are:
\begin{equation}
\mathsf{P}(\mathbf{r},t)=m\int
d\mathbf{v}\,\mathbf{V}\mathbf{V}f(\mathbf{r},
\mathbf{v},t)=p(\mathbf{r},t)\mathsf{I}+\int
d\mathbf{v}\,\mathsf{D}(\mathbf{ V})f(\mathbf{r},\mathbf{v},t),
\label{BD2.13}
\end{equation}
\begin{equation}
\mathbf{q}(\mathbf{r},t)=\frac{m}{2}\int
d\mathbf{v}\,V^{2}\mathbf{V}f( \mathbf{r},\mathbf{v},t)=\int
d\mathbf{v}\,\mathbf{S}(\mathbf{V})f(\mathbf{r} ,\mathbf{v},t).
\label{BD2.14}
\end{equation}
Here $\mathsf{I}$ is the unit tensor, and
\begin{equation}
\mathsf{D}(\mathbf{V})=m\left(
\mathbf{V}\mathbf{V}-\frac{1}{d}V^{2}\mathsf{I }\right) ,\quad
\mathbf{S}(\mathbf{V})=\left( \frac{m}{2}V^{2}-\frac{d+2}{2}
T\right) \mathbf{V}.  \label{BD2.15}
\end{equation}

The Boltzmann equation (\ref{BD2.1}) admits a special solution
$f_{HCS}$ describing the so-called homogeneous cooling state (HCS)
and having the scaling form \cite{BDGyS95}
\begin{equation}
f_{HCS}(\mathbf{v},t)=n_{H}v_{0}^{-d}(t)\chi
_{HCS}(\mathbf{c}),\quad \mathbf{c}=\frac{\mathbf{v}}{v_{0}(t)}\,,
\label{BD2.16}
\end{equation}
where
\begin{equation}
v_{0}(t)=\left[ \frac{2T_{HCS}(t)}{m}\right] ^{1/2}
\label{BD2.17}
\end{equation}
is the ``thermal velocity'' and $\chi _{HCS}(\mathbf{c})$ is an
isotropic function of $\mathbf{c}$. As indicated in Eq.\
(\ref{BD2.16}), all the time dependence of $f_{HCS}$ occurs
through the homogeneous temperature $ T_{HCS}(t)$, which obeys the
equation
\begin{equation}
\frac{\partial T_{HCS}(t)}{\partial t}+T_{HCS}(t)\zeta
_{HCS}(t)=0, \label{BD2.18}
\end{equation}
with
\begin{equation}
\zeta
_{HCS}(t)=\frac{(1-\alpha^{2})\pi^{\frac{d-1}{2}}\sigma^{d-1}
n_{H} v_{0}(t)}{ 2\Gamma \left( \frac{d+3}{2}\right) d}\,\int
d\mathbf{c}\,\int d
\mathbf{c}_{1}\chi_{HCS}(\mathbf{c})\chi_{HCS}(\mathbf{c}_{1})|\mathbf{c}-
\mathbf{c}_{1}|^{3}. \label{BD2.19}
\end{equation}
Substitution of the scaling form, Eq.\ (\ref{BD2.16}), into the
Boltzmann equation leads to
\begin{equation}
\frac{1}{2}\zeta _{HCS}\frac{\partial }{\partial \mathbf{v}}\cdot
\left( \mathbf{v}f_{HCS}\right) =J[f_{HCS}|f_{HCS}].
\label{BD2.20}
\end{equation}
The explicit time dependence of the HCS can be eliminated by
introducing a dimensionless time scale $s$ by
\begin{equation}
s=\int_{0}^{t}dt^{\prime }\,\frac{v_{0}(t^{\prime })}{\ell }\,,
\label{BD2.21}
\end{equation}
where $\ell \equiv \left( n_{H}\sigma ^{d-1}\right) ^{-1}$ is
proportional to the mean free path of the particles in the HCS. In
terms of the $s$ variable, Eq.\ (\ref{BD2.18}) becomes
\begin{equation}
\frac{\partial }{\partial s}T_{HCS}(s)=-\zeta _{0}T_{HCS}(s),
\label{BD2.22}
\end{equation}
with
\begin{equation}
\zeta _{0}=\frac{\ell \zeta _{HCS}}{v_{0}(t)}\,,  \label{BD2.23}
\end{equation}
that does not depend on time. Equation (\ref{BD2.22}) is the
differential form of the Haff's homogeneous cooling law
\cite{BDHa83}. Similarly, Eq.\ (\ref{BD2.20}) takes the form
\begin{equation}
\frac{1}{2}\zeta _{0}\frac{\partial }{\partial \mathbf{c}}\cdot
\left( \mathbf{c}\chi _{HCS}\right) =J^{\prime }[\chi _{HCS}|\chi
_{HCS}], \label{BD2.24}
\end{equation}
\begin{equation}
J^{\prime }\left[ \chi _{HCS}|\chi _{HCS}\right] =\int
d\mathbf{c}_{1}\,\bar{ T}(\mathbf{c},\mathbf{c}_{1})\chi
_{HCS}(\mathbf{c})\chi _{HCS}(\mathbf{c} _{1}),  \label{BD2.25}
\end{equation}
\begin{equation}
\bar{T}(\mathbf{c},\mathbf{c}_{1})=\int
d\hat{\boldsymbol{\sigma}}\Theta \left[
(\mathbf{c}-\mathbf{c}_{1})\cdot \hat{\boldsymbol{\sigma}}\right]
( \mathbf{c}-\mathbf{c}_{1})\cdot \hat{\boldsymbol{\sigma}}(\alpha
^{-2}b_{ \boldsymbol{\sigma}}^{-1}-1).  \label{BD2.26}
\end{equation}
The operator $b_{\boldsymbol{\sigma}}^{-1}$ in the last equation
is defined again by Eqs. (\ref{BD2.4}), but substituting
$(\mathbf{v},\mathbf{v}_{1})$ by $(\mathbf{c},\mathbf{c}_{1})$.
The form of the distribution function $\chi_{HCS}$ given by the
solution of Eq.\ (\ref{BD2.24}) has been extensively studied. A
summary of the results is given in Ref. \cite{BDvNyE01}.

\section{Linearized Boltzmann Equation}
\label{BDs2}Let us now consider a situation where the inelastic
gas is very close to the HCS, so that we can write
\begin{equation}
f(\mathbf{r},\mathbf{v},t)=f_{HCS}(\mathbf{v},t)+\delta
f(\mathbf{r},\mathbf{ v},t),\quad |\delta
f(\mathbf{r},\mathbf{v},t)|\ll f_{HCS}(\mathbf{v},t).
\label{BD2.27}
\end{equation}
Substitution of Eq.\ (\ref{BD2.27}) into the Boltzmann equation
(\ref{BD2.1} ), keeping only terms up to first order in $\delta f$
yields
\begin{equation}
\frac{\partial }{\partial t}\delta
f(\mathbf{r},\mathbf{v},t)+\mathbf{v} \cdot \nabla \delta
f(\mathbf{r},\mathbf{v},t)=K\delta f(\mathbf{r},\mathbf{v },t).
\label{BD2.28}
\end{equation}
The linearized Boltzmann collision operator is defined by
\begin{equation}
K\delta f(\mathbf{r},\mathbf{v},t)=J[f_{HCS}|\delta f]+J[\delta
f|f_{HCS}]. \label{BD2.29}
\end{equation}
Equation (\ref{BD2.28}) can be written in dimensionless form,
using the scaled velocity $\mathbf{c}$ defined in Eq.\
(\ref{BD2.16}), the reduced time $s$ introduced in Eq.\
(\ref{BD2.21}), and the dimensionless length scale
\begin{equation}
\mathbf{l}=\frac{\mathbf{r}}{\ell }.  \label{BD2.30}
\end{equation}
Then, the linearized inelastic Boltzmann equation becomes
\begin{equation}
\frac{\partial }{\partial s}\,\delta \chi +\mathbf{c}\cdot
\frac{\partial }{
\partial \mathbf{l}}\,\delta \chi =\Lambda \delta \chi ,  \label{BD2.31}
\end{equation}
where
\begin{equation}
\delta \chi (\mathbf{l},\mathbf{c},s)=n_{H}^{-1}v_{0}^{d}(t)\delta
f(\mathbf{ r},\mathbf{v},t)  \label{BD2.32}
\end{equation}
and
\begin{equation}
\Lambda \delta \chi =J^{\prime }[\chi _{HCS}|\delta \chi
]+J^{\prime }[\delta \chi |\chi _{HCS}]-\frac{\zeta
_{0}}{2}\frac{\partial }{\partial \mathbf{c}}\cdot \left(
\mathbf{c}\delta \chi \right) .  \label{BD2.33}
\end{equation}
In this representation, there is not any explicit dependence on
time due to the reference state, and all the stationary methods
known from the case of elastic collisions \cite{BDMc89} can now be
applied to the analysis of Eq.\ ( \ref{BD2.31}).

We define relative deviations of the hydrodynamic fields from
their values in the HCS by
\begin{equation}
\rho (\mathbf{l},s)\equiv \frac{\delta
n(\mathbf{r},t)}{n_{H}}=\int d\mathbf{ c}\,\delta \chi
(\mathbf{l},\mathbf{c},s),  \label{BD2.34}
\end{equation}
\begin{equation}
\boldsymbol{\omega}(\mathbf{l},s)\equiv \frac{\delta
\mathbf{u}(\mathbf{r},t) }{v_{0}(t)}=\int
d\mathbf{c}\,\mathbf{c}\delta \chi (\mathbf{l},\mathbf{c} ,s),
\label{BD2.35}
\end{equation}
\begin{equation}
\theta (\mathbf{l},s)\equiv \frac{\delta
T(\mathbf{r},t)}{T_{HCS}(t)}=\int d \mathbf{c}\,\left(
\frac{2c^{2}}{d}-1\right) \delta \chi (\mathbf{l},\mathbf{ c},s),
\label{BD2.36}
\end{equation}
where $\delta y(\mathbf{r},t)\equiv y(\mathbf{r},t)-y_{HCS}(t)$
denotes the deviation of the local value of the macroscopic
variable $y$ from its value in the HCS. Taking velocity moments in
the Boltzmann equation (\ref{BD2.31}), the linearized balance
equations for the hydrodynamic fields are obtained,
\begin{equation}
\frac{\partial }{\partial s}\rho (\mathbf{l},s)+\frac{\partial
}{\partial \mathbf{l}}\cdot \boldsymbol{\omega}(\mathbf{l},s)=0,
\label{BD2.37}
\end{equation}
\begin{equation}
\left( \frac{\partial }{\partial s}-\frac{\zeta _{0}}{2}\right)
\boldsymbol{\omega}(\mathbf{l},s)+\frac{\partial }{\partial
\mathbf{l}}\cdot \mathsf{\Pi }(\mathbf{l},s)=0,  \label{BD2.38}
\end{equation}
\begin{equation}
\frac{\partial }{\partial s}\theta
(\mathbf{l},s)+\frac{2}{d}\frac{\partial }{\partial
\mathbf{l}}\cdot \boldsymbol{\omega}(\mathbf{l},s)+\frac{2}{d}
\frac{\partial }{\partial \mathbf{l}}\cdot
\boldsymbol{\phi}(\mathbf{l} ,s)+\delta \zeta
_{0}(\mathbf{l},s)-\zeta _{0}\left[ \theta (\mathbf{l} ,s)+\rho
(\mathbf{l},s)\right] =0.  \label{BD2.39}
\end{equation}
The reduced linear pressure tensor $\mathsf{\Pi }$ and heat flux $
\boldsymbol{\phi}$ are defined as
\begin{equation}
\mathsf{\Pi }=\frac{1}{2}\left[ \theta (\mathbf{l},s)+\rho
(\mathbf{l},s) \right] \mathsf{I}+\int d\mathbf{c}\,\mathsf{\Delta
}(\mathbf{c})\delta \chi (\mathbf{l},\mathbf{c},s), \label{BD2.40}
\end{equation}
\begin{equation}
\boldsymbol{\phi}(\mathbf{l},s)=\int d\mathbf{c}\,\mathbf{\Sigma
}(\mathbf{c} )\delta \chi (\mathbf{l},\mathbf{c},s),
\label{BD2.41}
\end{equation}
where $\mathsf{\Delta}$ and $\boldsymbol{\phi}$ are the
dimensionless forms of $\mathsf{D}$ and $\mathbf{S}$,
\begin{equation}
\mathsf{\Delta
}(\mathbf{c})=\mathbf{c}\mathbf{c}-\frac{c^{2}}{d}\mathsf{I}
,\quad \boldsymbol{\Sigma}(\mathbf{c})=\left(
c^{2}-\frac{d+2}{2}\right) \mathbf{c},  \label{BD2.41a}
\end{equation}
The quantities $\mathsf{\Pi }$ and $\boldsymbol{\phi}$ are related
with the linearization of the fluxes given in Eq.\ (\ref{BD2.13})
and (\ref{BD2.14}) by
\begin{equation}
\mathsf{\Pi
}(\mathbf{l},s)=\frac{\mathsf{P}(\mathbf{r},t)}{2n_{H}T_{HCS}(t)}
-\frac{\mathsf{I}}{2},\quad
\boldsymbol{\phi}(\mathbf{l},s)=\frac{m^{1/2}\mathbf{q}(\mathbf{r},t)}{
2^{1/2}n_{H}T_{HCS}^{3/2}(t)}. \label{BD2.42}
\end{equation}
Finally, the term $\delta \zeta _{0}(\mathbf{l},s)$ is the
dimensionless linear deviation of the cooling term from its value
in the HCS,
\begin{equation}
\delta \zeta _{0}(\mathbf{l},s)=\frac{(1-\alpha ^{2})\pi
^{\frac{d-1}{2}}}{ \Gamma \left( \frac{d+3}{3}\right) d}\int
d\mathbf{c}\,\int d\mathbf{c} _{1}\,\chi _{HCS}(\mathbf{c})\delta
\chi (\mathbf{l},\mathbf{c},s)|\mathbf{c} -\mathbf{c}_{1}|^{3}.
\label{BD2.43}
\end{equation}

Of course, Eqs. (\ref{BD2.37})--(\ref{BD2.39}) can be derived also
by linearizing the nonlinear balance equations
(\ref{BD2.10})-(\ref{BD2.12}). They are the starting point for
deriving hydrodynamic equations, since they describe the evolution
of the hydrodynamic fields in the linear approximation. This
requires closing the equations, transforming the formal
expressions of the pressure tensor, the heat flux, and the cooling
rate into expressions in terms of the own hydrodynamic fields.
These are the so-called constitutive relations.

\section{Eigenvalue Problem}

\label{BDs3} In this Section we explore solutions to the
linearized Boltzmann equation and characterize the hydrodynamic
modes. This is done through a formal analysis of the eigenvalue
problem for the associated linear operator. Since the equation is
linear and the collision operator $ \Lambda $ does not change the
space variable, it is sufficient to consider a single Fourier
mode, i.e. we take
\begin{equation}
\delta \chi (\mathbf{l},\mathbf{c},s)= e^{i\mathbf{k}\cdot
\mathbf{ l}}\delta \widetilde{\chi }(\mathbf{k},\mathbf{c},s).
\label{BD3.01}
\end{equation}
The linearized Boltzmann equation, Eq.\ (\ref{BD2.31}), then
becomes
\begin{equation}
\frac{\partial }{\partial s}\,\delta \widetilde{\chi }=\left(
\Lambda -i \mathbf{k}\cdot \mathbf{c}\right) \delta
\widetilde{\chi },  \label{BD3.02}
\end{equation}
whose formal solution is
\begin{equation}
\delta \widetilde{\chi }(\mathbf{k},\mathbf{c},s)=e^{s\left(
\Lambda -i \mathbf{k}\cdot \mathbf{c}\right) }\delta
\widetilde{\chi }(\mathbf{k}, \mathbf{c},0).  \label{BD3.03}
\end{equation}
The above expression shows most clearly that the possible
excitations for the gas under these conditions of small initial
perturbations are determined by the spectrum of the linear
operator $\Lambda -i\mathbf{k}\cdot \mathbf{c}$. This suggests a
study of the eigenvalue problem
\begin{equation}
\left( \Lambda -i\mathbf{k}\cdot \mathbf{c}\right) \xi
_{i}(\mathbf{k}, \mathbf{c})=\lambda _{i}\left( \mathbf{k}\right)
\xi _{i}(\mathbf{k},\mathbf{ c}).  \label{BD3.04}
\end{equation}

The problem is posed in a Hilbert space of functions of
$\mathbf{c}$ with a scalar product given by
\begin{equation}
<g|h>=\int d\mathbf{c}\,\chi _{HCS}^{-1}(\mathbf{c})g^{\ast
}(\mathbf{c})h( \mathbf{c}),  \label{BD.041}
\end{equation}
where $g^{\ast }$ denotes the complex conjugate of $g$. Of
particular interest here are the eigenfunctions
$\lambda_{\beta}^{h}$ and eigenvalues
$\chi_{\beta}^{h}(\mathbf{c})$ associated with linear
hydrodynamics. These can be identified as follows. First, at
$\mathbf{k}=0$ they correspond to those eigenvalues of $\Lambda $
alone that coincide with the zeroth order in the gradients
eigenvalues of the linearized balance equations
(\ref{BD2.37})-(\ref{BD2.39}), which are easily seen to be $0$,
$\zeta_{0}/2$, and $-\zeta_{0}/2$, i.e.
\begin{equation}
\Lambda
\xi_{\beta}^{h}(\mathbf{c})=\lambda_{\beta}^{h}\xi_{\beta}^{h}(\mathbf{c}),
\quad \lambda _{\beta}^{h}=0,\pm \frac{\zeta _{0}}{2}\, ,
\label{BD3.05}
\end{equation}
where  $\xi_{\beta}^{h}(\mathbf{c})$ stands for
$\xi_{\beta}^{h}(\mathbf{k}=0,\mathbf{c})$. More generally, at
finite $\mathbf{k}$ the hydrodynamic modes are defined to be the
solutions of Eq.\ (\ref{BD3.04}) that are continuously connected
to those of Eq.\ (\ref{BD3.05}) as $k\rightarrow 0$. The
eigenvalue problem (\ref{BD3.05}) is solved below, so the
existence of the hydrodynamic modes is assured. The formal
solution (\ref{BD3.03}) then can be decomposed in the form
\begin{equation}
\delta \widetilde{\chi }(\mathbf{k},\mathbf{c},s)=\sum_{\beta}
\widetilde{a}_{\beta}^{h}(\mathbf{k},s)
\xi_{\beta}^{h}(\mathbf{k},\mathbf{c}) +\delta
\widetilde{\chi}^{m}(\mathbf{k},\mathbf{c},s).  \label{BD3.06}
\end{equation}
The first term arises from the hydrodynamic modes while the second
one represents all the other ``microscopic'' excitations. The
hydrodynamic coefficients
$\widetilde{a}_{\beta}^{h}(\mathbf{k},s)$ can be expressed as
\begin{equation}
\widetilde{a}_{\beta}^{h}(\mathbf{k},s)=e^{s\lambda
_{\beta}^{h}\left( \mathbf{k} \right)
}<\bar{\xi}_{\beta}^{h}(\mathbf{k},\mathbf{c})|\delta \chi
(\mathbf{k},
\mathbf{c},0)>=<\bar{\xi}_{\beta}^{h}(\mathbf{k},\mathbf{c})|\delta
\chi ( \mathbf{k},\mathbf{c},s)>.  \label{BD3.07}
\end{equation}
Since $\Lambda -i\mathbf{k}\cdot \mathbf{c}$ is not self-adjoint,
the functions $\bar{\xi}_{\beta}^{h}(\mathbf{k},\mathbf{c})$
differ from $\xi _{\beta}^{h}(\mathbf{k},\mathbf{c})$ but together
they must form a biorthogonal set. It will appear below that these
coefficients are closely related to the  hydrodynamic fields of
the linearized balance equations.

The expected context for a  hydrodynamic description is now clear.
For given initial conditions both hydrodynamic and microscopic
excitations contribute to (\ref{BD3.06}). However, for long
wavelengths (small $k$) it is expected that the microscopic
excitations decay on time scales short compared to those for the
hydrodynamic modes, leaving only the latter on long time scales.
The relevant separation of time scales is expected to occur after
a few collisions, so there is a large and interesting time scale
for the hydrodynamic description. Preliminary analysis based on
kinetic models for the Boltzmann equation
\cite{BDBMyD96,BDBDyS99,BDBMyR98} and also on low order matrix
representations of the linearized Boltzmann collision operator
\cite{BDDByL02}, suggest that the separation of modes assumed here
does occur, and hydrodynamics does dominate for wavelengths and
times long compared with the mean free path and the mean free
time, respectively. Furthermore, it appears from these models that
the cooling rate $\zeta _{0}$ remains smaller than the collision
frequency even for strong inelasticity, so that inclusion of the
non-conserved energy among the hydrodynamic variables seems
justified.

\subsection*{Hydrodynamic modes for $k=0$}
\label{BDs3a}We consider now the eigenproblem associated with the
linearized homogeneous Boltzmann operator,
\begin{equation}
\Lambda \xi _{i}(\mathbf{c})=\lambda _{i}\xi _{i}(\mathbf{c}).
\label{BD3.1}
\end{equation}
Of course, finding all the solutions of this equation is an
impossible task, even in the elastic limit. Nevertheless, it is
quite easy to obtain some particular solutions, which will turn
out to be the relevant ones in the hydrodynamic regime. We begin
by constructing the function
\begin{equation}
F(\mathbf{c},\rho ,\boldsymbol{\omega},\gamma )\equiv \rho \chi
_{HCS}(\gamma \mathbf{C}),  \label{BD3.2}
\end{equation}
where $\mathbf{C}=\mathbf{c}-\boldsymbol{\omega}$. Although $F$
has the functional form of a ``local'' HCS distribution, the
parameters $\rho $, $ \boldsymbol{\omega}$, and $\gamma $ do not
need to have any particular meaning here, playing just the role of
auxiliary parameters. From Eq.\ (\ref {BD2.24}) it is obtained
that $F$ verifies
\begin{equation}
\frac{\rho }{2}\zeta _{0}\frac{\partial }{\partial
\mathbf{c}}\cdot \left( \mathbf{C}F\right) =\gamma
^{d+1}\mathit{J}^{\prime}\left[ F|F\right] .  \label{BD3.3}
\end{equation}
Then, taking derivatives with respect to $\rho $,
$\boldsymbol{\omega}$, and $\gamma $ in the above equation and,
afterwards, the limit $\rho =\gamma =1$ , $\boldsymbol{\omega}=0$,
leads to the set of equations
\begin{equation}
\Lambda \psi _{1}(\mathbf{c})=-\frac{1}{2}\zeta _{0}\psi
_{3}(\mathbf{c} ),\quad \Lambda
\boldsymbol{\psi}_{2}(\mathbf{c})=\frac{1}{2}\zeta _{0}
\boldsymbol{\psi}_{2}(\mathbf{c}),\quad \Lambda \psi
_{3}(\mathbf{c})=-\frac{ 1}{2}\zeta _{0}\psi _{3}(\mathbf{c}),
\label{BD3.4}
\end{equation}
where
\begin{equation}
\psi _{1}(\mathbf{c})\equiv \left( \frac{\partial F}{\partial \rho
}\right) _{0}=\chi _{HCS}(\mathbf{c}),  \label{BD3.5}
\end{equation}
\begin{equation}
\boldsymbol{\psi}_{2}(\mathbf{c})\equiv \left( \frac{\partial
F}{\partial \boldsymbol{\omega}}\right) _{0}=-\frac{\partial \chi
_{HCS}(\mathbf{c})}{
\partial \mathbf{c}},  \label{BD3.6}
\end{equation}
\begin{equation}
\psi _{3}(\mathbf{c})\equiv -\left( \frac{\partial F}{\partial
\gamma } \right) _{0}-\chi _{HCS}(\mathbf{c})d=-\frac{\partial
}{\partial \mathbf{c}} \cdot \left[ \mathbf{c}\chi
_{HCS}(\mathbf{c})\right] .  \label{BD3.7}
\end{equation}
Here the zero subindex indicates that the derivatives are
evaluated at $\rho =\gamma =1,\boldsymbol{\omega}=0$. From Eqs.
(\ref{BD3.4}) it is easily seen that the set of functions
\begin{equation}
\left\{ \xi _{\beta }(\mathbf{c})\right\}  \equiv \left\{
\psi_{1}(\mathbf{c}
)-\psi_{3}(\mathbf{c}),\boldsymbol{\psi}_{2}(\mathbf{c}),
\psi_{3}(\mathbf{c})\right\}   \label{BD3.8}
\end{equation}
are solutions of Eq.\ (\ref{BD3.1}) corresponding to the
eigenvalues
\begin{equation}
\lambda _{1}=0,\quad \lambda _{2}=\frac{\zeta _{0}}{2}\,,\quad
\lambda _{3}=- \frac{\zeta _{0}}{2}\,,  \label{BD3.9}
\end{equation}
respectively. The eigenvalue $\lambda _{2}$ is, therefore,
$d$-fold degenerate. This confirms (\ref{BD3.05}) above,
indicating points in the spectrum at $k=0$ corresponding to the
linear balance equations. Hence, these are the hydrodynamic modes
at longest wavelengths, $\xi_{\beta}^{h}(\mathbf{c})$.

The eigenfunctions $\xi _{\beta }^{h}(\mathbf{c})$ can be
interpreted as follows. Consider a local form of the HCS
distribution, $f_{HCS}^{(l)}$, whereby the hydrodynamics fields
are replaced by their actual space and time dependent values in a
general nonequilibrium state,
\begin{equation}
f_{HCS}^{(l)}(\mathbf{r},\mathbf{v},t)=n(\mathbf{r},t)v_{0}^{(l)-d}(t)
\chi_{HCS}(\mathbf{C}),  \label{BD3.9a}
\end{equation}
where
\begin{equation}
\mathbf{C}=\frac{\mathbf{v}-\delta
\mathbf{u}(\mathbf{r},t)}{v_{0}^{(l)}(t)} ,\quad
v_{0}^{(l)}(t)=\left[ \frac{2T(\mathbf{r},t)}{m}\right] ^{1/2}.
\label{BD3.9b}
\end{equation}
This is the analogous to the local equilibrium distribution for
elastic collisions. It represents a state whose averages values of
$1$, $\mathbf{v}$ , and $v^{2}$ are those of the nonequilibrium
state. This seems a reasonable choice to describe an
experimentally prepared state. The linearization of such an state
around the HCS gives
\begin{equation}
f_{HCS}^{(l)}(\mathbf{r},\mathbf{v},t)\rightarrow
f_{HCS}(\mathbf{v} ,t)+n_{H}v_{0}^{-d}(t)\delta \chi
^{(l)}(\mathbf{c},\mathbf{l},s), \label{BD3.9c}
\end{equation}
\begin{equation}
\delta \chi ^{(l)}(\mathbf{c},\mathbf{l},s)=\left[
\xi_{1}^{h}(\mathbf{c})+\xi_{3}^{h}(\mathbf{c})\right] \rho
(\mathbf{l},s) +\boldsymbol{\xi}_{2}^{h}(\mathbf{c})\cdot
\boldsymbol{\omega}(\mathbf{l},s)+\frac{1}{2}\xi
_{3}^{h}(\mathbf{c} )\theta (\mathbf{l},s).  \label{BD3.9d}
\end{equation}
Thus, the eigenfunctions $\xi_{\beta}^{h}(\mathbf{c})$ describe
the linear deviation of the local HCS, appearing then as the
natural perturbations for a macroscopic preparation of the initial
nonequilibrium state. This will be discussed further below.

It is easy to show that the functions
$\{\xi_{\beta}^{h}(\mathbf{c})\}$ are linearly independent so that
they define a $d+2$ dimensional subspace of the Hilbert space.
Although the functions $\psi _{\beta }(\mathbf{c})$ are orthogonal
with the definition of scalar product in Eq.\ (\ref{BD.041}), the
eigenfunctions $\xi _{\beta}^{h}(\mathbf{c})$ are not, as a
consequence of the operator $\Lambda $ being non-Hermitian. This
leads, in principle, to consider the left eigenvalue problem and,
therefore, to study the adjoint operator $\Lambda ^{+}$ defined by
\begin{equation}
<g|\Lambda h>^{\ast }=<h|\Lambda ^{+}g>,  \label{BD3.11}
\end{equation}
for arbitrary $g(\mathbf{c})$ and $h(\mathbf{c})$. Using the
property given in Eq.\ (\ref{BD2.5}) it is found
\begin{eqnarray}
\label{BD3.12} \Lambda ^{+}g ({\bf c}) &=&\int
d\mathbf{c}_{1}\,\chi_{HCS}(\mathbf{c})\chi_{HCS}(
\mathbf{c}_{1})T(\mathbf{c},\mathbf{c}_{1})\left[
\chi_{HCS}^{-1}(\mathbf{c}
)g(\mathbf{c})+\chi_{HCS}^{-1}(\mathbf{c}_{1})g(\mathbf{c}_{1})\right]
 \nonumber \\
 &&+\frac{\zeta _{0}}{2}\chi
_{HCS}(\mathbf{c})\mathbf{c}\cdot \frac{\partial }{\partial
\mathbf{c}}\left[ \chi
_{HCS}^{-1}(\mathbf{c})g(\mathbf{c})\right].
\end{eqnarray}
The eigenfunctions of $\Lambda ^{+}$ would provide the left
eigenfunctions of $\Lambda $. We have not been able to find all
the eigenfunctions corresponding to the eigenvalues in Eq.\
(\ref{BD3.9}). Nevertheless, for our purposes here, and also for
many other applications, it will suffice to identify a set of
functions $\left\{ \bar{\xi}_{\beta }^{h}(\mathbf{c}),\beta
=1,2,3\right\} $ verifying the biorthonormality condition
\begin{equation}
<\bar{\xi}_{\beta }^{h}|\xi _{\beta ^{\prime }}^{h}>=\delta
_{\beta ,\beta^{\prime }},  \label{BD3.13}
\end{equation}
$\beta ,\beta ^{\prime }=1,2,3$. A convenient choice is given by
\begin{equation}
\bar{\xi}_{1}^{h}(\mathbf{c})=\chi _{HCS}(\mathbf{c}),\quad
\bar{\boldsymbol{\xi} }_{2}^{h}(\mathbf{c})=\mathbf{c}\chi
_{HCS}(\mathbf{c}),\quad \bar{\xi}_{3}^{h}=\left(
\frac{c^{2}}{d}+\frac{1}{2}\right) \chi _{HCS}(\mathbf{c}).
\label{BD3.14}
\end{equation}
The above functions span a dual subspace of that spanned by the
right eigenfunctions $\{\xi _{\beta }^{h}(\mathbf{c})\}$. The fact
that they are linear combinations of $1$, $\mathbf{c}$, and
$c^{2}$ is closely related with the existence and approach to a
hydrodynamic description, as it will be shown in the next Section.
Let us stress that while $\bar{\xi}_{1}^{h}(\mathbf{c} )$ and
$\bar{\boldsymbol{\xi}}_{2}^{h}(\mathbf{c})$ are eigenfunctions of
$ \Lambda ^{+}$ corresponding to the eigenvalues $0$ and $\zeta
_{0}/2$, respectively, the function
$\bar{\xi}_{3}^{h}(\mathbf{c})$ is not an eigenfunction of
$\Lambda ^{+}$. The important point is that for any linear
combination
\begin{equation}
g(\mathbf{c})=\sum_{\beta =1}^{3}a_{\beta } \xi
_{\beta}^{h}(\mathbf{c}), \label{BD3.15}
\end{equation}
the coefficients $a_{\beta }$ are given by
\begin{equation}
a_{\beta }=<\bar{\xi}_{\beta }^{h}|g>=\int d\mathbf{c}\,\chi
_{HCS}^{-1}(\mathbf{ c})\bar{\xi}_{\beta
}^{h}(\mathbf{c})g(\mathbf{c}).  \label{BD3.16}
\end{equation}

It is instructive to consider what happens in the elastic limit
$\alpha \rightarrow 1$. Then $\chi _{HCS}(\mathbf{c})$ becomes the
Maxwellian $\chi _{\mathrm{MB}}(\mathbf{c})=\pi ^{-d/2}\exp
(-c^{2})$ and the right eigenfunctions are given by
\begin{equation}
\left\{ \left[ 2c^{2}-(d+1)\right] \chi
_{\mathrm{MB}}(\mathbf{c}),2\mathbf{c }\chi
_{\mathrm{MB}}(\mathbf{c}),(2c^{2}-d)\chi
_{\mathrm{MB}}(\mathbf{c} )\right\} .  \label{BD3.17}
\end{equation}
Of course, these functions are not orthogonal and differ from
those usually employed in the linear analysis of the elastic
Boltzmann equation \cite {BDRydL77}. This is an important point,
since when solving the eigenvalue problem associated with the
inelastic homogeneous Boltzmann equation by means of a
perturbation calculus around the elastic limit, the first step in
the perturbation scheme is to remove the degeneracy of the elastic
zeroth eigenvalue by exactly solving the perturbed eigenvalue
problem within the subspace spanned by the degenerate
eigenfunctions. The relevant conclusion is that the functions
$\xi_{\beta}^{h}(\mathbf{c})$ appear in a natural way even when
considering the small inelasticity limit.

\section{Navier-Stokes and Green-Kubo expressions}

\label{BDs4} In the last section the hydrodynamic modes were
defined and calculated at asymptotically long wavelengths. It
remains to identify the hydrodynamic coefficients in Eq.\
(\ref{BD3.06}) as well as the hydrodynamic modes at finite $k$.
Consider first the coefficients defined by Eq.\ (\ref{BD3.07} )
and evaluate the biorthogonal function
$\bar{\xi}_{\beta}^{h}(\mathbf{k}, \mathbf{c})$ to leading order
in $k$, so that
\begin{equation}
\widetilde{a}_{\beta}^{h}(\mathbf{k},s)\rightarrow
<\bar{\xi}_{\beta}^{h}(\mathbf{c} )|\delta \chi
(\mathbf{k},\mathbf{c},s)>.  \label{BD4.01}
\end{equation}
Use of the expressions given in Eq.\ (\ref{BD3.14}) shows that, in
coordinate representation, these are just the hydrodynamic fields
of the linear balance equations
\begin{equation}
\left\{ a_{\beta}^{h}(\mathbf{l},s)\right\} \rightarrow \left\{
\rho
(\mathbf{l},s),\boldsymbol{\omega}(\mathbf{l},s),\frac{1}{2}\theta
(\mathbf{l},s)+\rho ( \mathbf{l},s)\right\} .  \label{BD4.02}
\end{equation}
This completes the connection between the hydrodynamic modes and
the dynamics determined from the linear balance equations. They
are the same for long wavelengths and small perturbations, as
expected.

\subsection{Navier-Stokes Approximation}

\label{BDs4a}To obtain the hydrodynamic eigenvalues
$\lambda_{\beta}^{h}\left( \mathbf{k}\right) $ to Navier-Stokes
($k^{2}$) order, the eigenvalue problem (\ref{BD3.04}) can be
solved treating $i\mathbf{c}\cdot \mathbf{k}$ as a small
perturbation of the operator $\Lambda $
\cite{BDDByRunpublished03}. This method implicitly restricts $k$
to be small relative to the deviation of the restitution
coefficient $\alpha $ from unity, the elastic limit. A more
uniform description, without restrictions on $k$ relative to
$\alpha $, is obtained by calculating the fluxes in the balance
equations as functions of $\widetilde{a}_{\beta}^{h}(
\mathbf{k},s)$, to first order in $k$. The resulting Navier-Stokes
hydrodynamic equations can then be used to determine $\lambda
_{i}^{h}\left( \mathbf{k}\right)$. This is the approach used here.
The first step is accomplished by using the identity
\begin{equation}
e^{(A+B)s}=e^{As}+\int_{0}^{s}ds^{\prime
}\,e^{As^{\prime}}Be^{(A+B)(s-s^{\prime })},  \label{BD4.2a}
\end{equation}
valid for arbitrary operators $A$ and $B$, in the formal solution
(\ref {BD3.03}) to get
\begin{equation}
\delta \widetilde{\chi }(\mathbf{k},\mathbf{c},s)=e^{s\Lambda
}\delta \widetilde{\chi
}(\mathbf{k},\mathbf{c},0)-\int_{0}^{s}ds^{\prime }\,e^{s^{\prime
}\Lambda }i\mathbf{k}\cdot \mathbf{c}e^{-s^{\prime }\left( \Lambda
-i\mathbf{k}\cdot \mathbf{c}\right) }\delta \widetilde{\chi }(
\mathbf{k},\mathbf{c},s).  \label{BD4.2}
\end{equation}
Next, substitute the representation (\ref{BD3.06}) to identify the
dependence on $\widetilde{a}_{\beta}^{h}(\mathbf{k},s)$,
\begin{eqnarray}
\label{BD4.202}
 \delta \widetilde{\chi }(\mathbf{k},\mathbf{c},s)
&=&e^{s\Lambda }\delta \widetilde{\chi
}(\mathbf{k},\mathbf{c},0)-\int_{0}^{s}ds^{\prime }\,e^{s^{\prime
}\Lambda }i\mathbf{k}\cdot \mathbf{c}e^{-s^{\prime }\left( \Lambda
-i\mathbf{k}\cdot \mathbf{c}\right) } \notag \\
&& \times \left[ \sum_{\beta}
\widetilde{a}_{\beta}^{h}(\mathbf{k},s)
\xi_{\beta}^{h}(\mathbf{k},\mathbf{c})
 +\delta \widetilde{\chi
}^{m}(\mathbf{k},\mathbf{c},s)\right]
\notag \\
&\rightarrow &e^{s\Lambda }\delta \widetilde{\chi
}(\mathbf{k},\mathbf{c} ,0)-\sum_{\beta}\left[
\int_{0}^{s}ds^{\prime }\,e^{s^{\prime }\Lambda }\mathbf{c}
e^{-s^{\prime } \Lambda }\xi _{\beta}^{h}(\mathbf{c})\right] \cdot
i\mathbf{k}
\widetilde{a}_{\beta}^{h}(\mathbf{k},s)  \notag \\
&=&e^{s\Lambda }\delta \widetilde{\chi
}(\mathbf{k},\mathbf{c},0)-\sum_{\beta} \left[
\int_{0}^{s}ds^{\prime }\,e^{s^{\prime }\left(
\Lambda-\lambda_{\beta}^{h}\right) }\mathbf{c}
\xi_{\beta}^{h}(\mathbf{c})\right] \cdot i\mathbf{k}
\widetilde{a}_{\beta}^{h}(\mathbf{k},s). \nonumber \\
\end{eqnarray}

In the second transformation, it has been assumed that $k$ is
sufficiently small so that only leading order terms need to be
retained, and that $s$ is sufficiently large to neglect the
microscopic excitations relative to the hydrodynamic modes
contributions. Moreover, in the last equality use has been made of
the fact that $\xi_{\beta}^{h}$ is an eigenfunction of $\Lambda$.
In coordinate representation this result becomes more explicitly
\begin{equation}
\delta \chi (\mathbf{l},\mathbf{c},s)\rightarrow e^{s\Lambda
}\delta \chi ( \mathbf{l},\mathbf{c},0)-\sum_{\beta
=1}^{3}\mathbf{F}_{\beta }(\mathbf{c} ,s)\cdot \frac{\partial
}{\partial \mathbf{l}}\nu _{\beta }(\mathbf{l},s), \label{BD4.9}
\end{equation}
where
\begin{equation}
\left\{ \nu _{\beta }(\mathbf{l},s)\right\} \equiv \left\{ \rho
(\mathbf{l} ,s),\boldsymbol{\omega}(\mathbf{l},s),\theta
(\mathbf{l},s)\right\} \label{BD4.10}
\end{equation}
and
\begin{equation}
\mathbf{F}_{1}(\mathbf{c},s)=\int_{0}^{s}ds^{\prime }e^{s^{\prime
}\Lambda }\xi
_{1}(\mathbf{c})\mathbf{c}+2\mathbf{F}_{3}(\mathbf{c},s),
\label{BD4.11}
\end{equation}
\begin{equation}
F_{2,ij}(\mathbf{c},s)=\int_{0}^{s}ds^{\prime }e^{s^{\prime
}\left( \Lambda - \frac{\zeta _{0}}{2}\right) }\xi
_{2,i}(\mathbf{c})c_{j},  \label{BD4.12}
\end{equation}
\begin{equation}
\mathbf{F}_{3}(\mathbf{c},s)=\frac{1}{2}\int_{0}^{s}ds^{\prime
}\,e^{s^{\prime }\left( \Lambda +\frac{\zeta _{0}}{2}\right) }\xi
_{3}( \mathbf{c})\mathbf{c}.  \label{BD4.13}
\end{equation}

The pressure tensor and the heat flux up to first order in the
gradients of the fields are now obtained by substituting Eq.\
(\ref{BD4.9}) into Eqs.\ (\ref{BD2.40}) and (\ref{BD2.41}). The
term involving the initial condition $ \delta \xi
(\mathbf{l},\mathbf{c},0)$ does not contribute to these quantities
since, in the long time limit, it is given by a linear combination
of the functions $\chi _{\beta }^{h}(\mathbf{c})$, that are all
orthogonal to both $\mathsf{\Delta }(\mathbf{c})\chi
_{HCS}(\mathbf{c})$ and $ \boldsymbol{\Sigma}(\mathbf{c})\chi
_{HCS}(\mathbf{c})$. Taking into account the symmetry properties
of the system, the resulting expressions can be written in the
form
\begin{eqnarray}
\Pi _{ij}(\mathbf{l},s) &=&\frac{1}{2}\left[ \theta
(\mathbf{l},s)+\rho (
\mathbf{l},s)\right] \delta _{ij}  \notag \\
&&-\tilde{\eta}\left[ \frac{\partial }{\partial l_{i}}\omega
_{j}(\mathbf{l} ,s)+\frac{\partial }{\partial l_{j}}\omega
_{i}(\mathbf{l},s)-\frac{2}{d} \frac{\partial }{\partial
\mathbf{l}}\cdot \boldsymbol{\omega}(\mathbf{l} ,s)\delta
_{ij}\right] ,  \label{BD4.14}
\end{eqnarray}
\begin{equation}
\boldsymbol{\phi}(\mathbf{l},s)=-\tilde{\kappa}\frac{\partial
}{\partial \mathbf{l}}\theta
(\mathbf{l},s)-\tilde{\mu}\frac{\partial }{\partial
\mathbf{l}}\rho (\mathbf{l},s).  \label{BD4.15}
\end{equation}
Equation (\ref{BD4.14}) is the expected Navier-Stokes expression
for the pressure tensor, involving the shear viscosity coefficient
$\tilde{\eta}$, but Eq.\ (\ref{BD4.15}) contains, besides the
usual Fourier law characterized by the heat conductivity
$\tilde{\kappa}$, an additional contribution proportional to the
density gradient and with an associated transport coefficient
$\tilde{\mu}$. This latter term has no analogue in elastic gases.

The expressions of the (time-dependent) transport coefficients are
\begin{equation}
\tilde{\eta}(s)=\frac{1}{d^{2}+d-2}\sum_{i,j}^{d}\int
d\mathbf{c}\,\Delta _{ij}(\mathbf{c})F_{2,ij}(\mathbf{c},s),
\label{BD4.16}
\end{equation}
\begin{equation}
\tilde{\kappa}(s)=\frac{1}{d}\int
d\mathbf{c}\,\boldsymbol{\Sigma}(\mathbf{c} )\cdot
\mathbf{F}_{3}(\mathbf{c},s),  \label{BD4.17}
\end{equation}
\begin{equation}
\tilde{\mu}=\frac{1}{d}\int
d\mathbf{c}\boldsymbol{\Sigma}(\mathbf{c})\cdot
\mathbf{F}_{1}(\mathbf{c},s).  \label{BD4.18}
\end{equation}
As already mentioned, the functions $\Delta _{ij}(\mathbf{c})\chi
_{HCS}( \mathbf{c})$ and $\Sigma _{i}(\mathbf{c})\chi
_{HCS}(\mathbf{c})$ are orthogonal to the subspace spanned by the
right eigenfunctions $\xi _{\beta }(\mathbf{c})$. Therefore, the
decay of the ``correlation functions'' in Eqs.\
({\ref{BD4.16})--(\ref{BD4.18}) lies outside the spectrum of the
slowest modes, and the $s$-time integral in the expression of the
functions $F_{\beta }$ converges. Consequently, it is possible to
set $s\rightarrow \infty $ for times scales long compared with the
microscopic times. In this way, the reduced transport coefficients
become time-independent. Returning to the original variables, the
above expressions for the hydrodynamic fluxes are equivalent to
\begin{eqnarray}
P_{ij}(\mathbf{r},t) &=&n_{H}T_{HCS}(t)\left[ 1+\frac{\delta
T(\mathbf{r},t) }{T_{HCS}(t)}+\frac{\delta
n(\mathbf{r},t)}{n_{H}}\right] \delta _{ij}
\notag  \label{BD4.19} \\
&&-\eta \left[ \frac{\partial }{\partial r_{i}}\delta
u_{j}+\frac{\partial }{
\partial r_{j}}\delta u_{i}-\frac{2}{d}(\boldsymbol{\nabla}\cdot \delta
\mathbf{u})\delta _{ij}\right] ,
\end{eqnarray}
\begin{equation}
\mathbf{q}(\mathbf{r},t)=-\kappa \boldsymbol{\nabla}\delta
T(\mathbf{r} ,t)-\mu \boldsymbol{\nabla}\delta n(\mathbf{r},t),
\label{BD4.20}
\end{equation}
with the transport coefficients given by
\begin{equation}
\eta =n_{H}m\ell v_{0}(t)\tilde{\eta},  \label{BD4.21}
\end{equation}
\begin{equation}
\kappa =n_{H}\ell v_{0}(t)\tilde{\kappa},  \label{BD4.22}
\end{equation}
\begin{equation}
\mu =\frac{m\ell v_{0}^{3}(t)}{2}\tilde{\mu}.  \label{BD4.23}
\end{equation}
}

\subsection{Green-Kubo Relations}

\label{BDs4b}{It is possible to write Eqs.
(\ref{BD4.16})--(\ref{BD4.18}) in the form of Green-Kubo
relations. Define an operator $\tilde{\Lambda}$ by
\begin{equation}
\left[ \int d\mathbf{c}\,g^{\ast }(\mathbf{c})\Lambda
h(\mathbf{c})\right] ^{\ast }=\int d\mathbf{c}\,h^{\ast
}(\mathbf{c})\tilde{\Lambda}g(\mathbf{c}), \label{BD4.24}
\end{equation}
for arbitrary $h$ and $g$. Comparison with Eq.\ (\ref{BD3.11})
yields
\begin{eqnarray}
\tilde{\Lambda}g(\mathbf{c}) &=&\chi
_{HCS}^{-1}(\mathbf{c})\Lambda ^{+}
\left[ g(\mathbf{c})\chi _{HCS}(\mathbf{c})\right]   \notag \\
&=&\int d\mathbf{c}_{1}\chi
_{HCS}(\mathbf{c}_{1})T(\mathbf{c},\mathbf{c}
_{1})[g(\mathbf{c})+g(\mathbf{c}_{1})]+\frac{\zeta
_{0}}{2}\mathbf{c}\cdot \frac{\partial }{\partial
\mathbf{c}}g(\mathbf{c}).  \label{BD4.25}
\end{eqnarray}
Also define ``correlation functions'' in the HCS by
\begin{equation}
<gh>\equiv \int d\mathbf{c}\,\chi
_{HCS}(\mathbf{c})g(\mathbf{c})h(\mathbf{c} ).  \label{BD4.26}
\end{equation}
Then, the expressions for the transport coefficients can be
rewritten as
\begin{equation}
\tilde{\eta}=\frac{1}{d^{2}+d-2}\sum_{i,j}^{d}\int_{0}^{\infty
}ds\,<\Delta _{ij}(s)\Phi _{2,ij}>e^{-s\zeta _{0}/2},
\label{BD4.27}
\end{equation}
\begin{equation}
\tilde{\kappa}=\frac{1}{d}\int_{0}^{\infty
}ds\,<\boldsymbol{\Sigma}(s)\cdot \boldsymbol{\Phi}_{3}>e^{s\zeta
_{0}/2},  \label{BD4.28}
\end{equation}
\begin{equation}
\tilde{\mu}=2\tilde{\kappa}+\frac{1}{d}\int_{0}^{\infty }ds\,<
\boldsymbol{\Sigma}(s)\cdot \boldsymbol{\Phi}_{1}>,
\label{BD4.29}
\end{equation}
with the definitions
\begin{equation}
\boldsymbol{\Phi}_{1}\equiv
\mathbf{c}-2\boldsymbol{\Phi}_{3}(\mathbf{c}), \label{BD4.30}
\end{equation}
\begin{equation}
\Phi _{2,ij}(\mathbf{c})=-c_{j}\frac{\partial \ln \chi
_{HCS}(\mathbf{c})}{
\partial c_{i}},  \label{BD4.31}
\end{equation}
\begin{equation}
\boldsymbol{\Phi}_{3}(\mathbf{c})=-\frac{1}{2}\mathbf{c}\left[
d+\mathbf{c} \cdot \frac{\partial \ln \chi
_{HCS}(\mathbf{c})}{\partial \mathbf{c}}\right] .  \label{BD4.32}
\end{equation}
The time dependence of the dynamical variables is given by
\begin{equation}
g(\mathbf{c},s)=e^{s\tilde{\Lambda}}g(\mathbf{c}).  \label{BD4.33}
\end{equation}
Equations (\ref{BD4.27})--(\ref{BD4.29}) are the Green-Kubo
formulae for a dilute granular gas described by the Boltzmann
equation. They express the transport coefficients as integrals of
time-correlations functions in the HCS. Although they are not
identical to those derived in Ref. \cite{BDDyB02} by a different
method, it is easy to show they are equivalent, since they differ
in terms giving vanishing contributions.

Let us also compare them with the expressions derived from the
nonlinear Boltzmann equation by the Chapmann-Enskog method
\cite{BDBDKyS98,BDByC01}. Consider the reduced shear viscosity
coefficient. It is given by Eq. (\ref{BD4.16}) with $F_{2,ij}$
defined in Eq.\ (\ref{BD4.12}). In the limit of large $s$, we have
seen that the integral in the latter expression can be extended up
to infinity. Performing then the integral, it is obtained
\begin{equation}
\tilde{\eta}=-\frac{1}{d^{2}+d-2}\sum_{i,j}^{d}\int
d\mathbf{c}\,\Delta _{ij}(\mathbf{c})\left( \Lambda -\frac{\zeta
_{0}}{2}\right) ^{-1}\xi _{2,i}( \mathbf{c})c_{j},  \label{BD4.34}
\end{equation}
or, equivalently,
\begin{equation}
\tilde{\eta}=\frac{1}{d^{2}+d-2}\sum_{i,j}^{d}\int
d\mathbf{c}\,\Delta _{ij}(
\mathbf{c})\mathcal{C}_{ij}(\mathbf{c}),  \label{BD4.35}
\end{equation}
where $\mathcal{C}_{ij}$ is the solution of the integral equation
\begin{equation}
\left( \Lambda -\frac{\zeta _{0}}{2}\right)
\mathcal{C}_{ij}(\mathbf{c} )=-\xi _{2,i}(\mathbf{c})c_{j},
\label{BD4.36}
\end{equation}
being orthogonal to the subspace spanned by the lowest order right
eigenfunctions of $\Lambda $. Equations (\ref{BD4.35}) and
(\ref{BD4.36}) are equivalent to those derived by the
Chapman-Enskog procedure in Refs. \cite{BDBDKyS98,BDByC01}, where
the explicit expressions of the transport coefficients in the
first Sonine approximation is obtained. A similar analysis leading
to the same conclusion can be carried out for the other transport
coefficients. The only conceptually relevant difference is in the
reference state. In the present linearized analysis the transport
coefficients are defined in the HCS, whereas in the
Chapmann-Enskog method they are defined in a \emph{local} HCS,
characterized by the parameters $n( \mathbf{r},t)$,
$\mathbf{u}(\mathbf{r},t)$ and $T(\mathbf{r},t)$. As a
consequence, these are the fields appearing in the scaling in Eqs.
(\ref {BD4.21})--(\ref{BD4.23}).

It is also instructive to consider the elastic limit $\alpha
\rightarrow 1$ of the Green-Kubo expressions derived here. In that
case, we already know that the HCS distribution becomes the
Maxwellian $\chi _{MB}$, and $\zeta_{0}\rightarrow 0$. Then, the
functions defined in Eqs.\ (\ref{BD4.30})--(\ref{BD4.32}) become
\begin{equation}
\boldsymbol{\Phi}_{1}(\mathbf{c})\rightarrow
-2\boldsymbol{\Sigma}(\mathbf{c} ),\quad \Phi
_{2,ij}(\mathbf{c})\rightarrow 2c_{i}c_{j},\quad
\boldsymbol{\Phi}_{3}(\mathbf{c})\rightarrow \left(
c^{2}-\frac{d}{2}\right) \mathbf{c}.  \label{BD4.37}
\end{equation}
Therefore, the transport coefficient $\tilde{\mu}$, given by Eq.\
(\ref {BD4.29}) vanishes in the elastic limit, as expected, and
\begin{eqnarray}
\eta \rightarrow \eta _{0} &=&\frac{2nm\ell v_{0}(t)}{d^{2}+d-2}
\sum_{i,j}^{d}\int_{0}^{\infty }ds\,<\Delta _{ij}(s)c_{i}c_{j}>  \notag \\
&=&\frac{2nm\ell
v_{0}(t)}{d^{2}+d-2}\sum_{i,j}^{d}\int_{0}^{\infty }ds\,<\Delta
_{ij}(s)\Delta _{ij}>.  \label{BD4.38}
\end{eqnarray}
\begin{eqnarray}
\kappa \rightarrow \kappa _{0} &=&\frac{n\ell
v_{0}(t)}{d}\int_{0}^{\infty }ds\,<\boldsymbol{\Sigma}(s)\cdot
\left( c^{2}-\frac{d}{2}\right) \mathbf{c}>
\notag \\
&=&\frac{n\ell v_{0}(t)}{d}\int_{0}^{\infty
}ds\,<\boldsymbol{\Sigma} (s)\cdot \boldsymbol{\Sigma}>.
\label{BD4.39}
\end{eqnarray}
The last equalities in the above equations follow from the
symmetry of the time correlation functions. For elastic systems,
the temperature is constant to lowest order in the gradients of
the hydrodynamic fields and, therefore, in the integrands
appearing in the above expressions the definition of the scale
$s$, given by Eq. (\ref{BD2.21}), reduces to $s=v_{0}t/\ell $,
i.e. it is simply proportional to the actual time. Then, Eqs.\
(\ref{BD4.38}) and ( \ref{BD4.39}) are the low density limit of
the usual Green-Kubo expressions for the transport coefficients in
terms of the autocorrelation functions of the microscopic fluxes
of the relevant densities \cite{BDMc89}.

\section{Discussion}

The primary results of the above analysis have been the
identification of hydrodynamic modes in the spectrum of the
linearized Boltzmann operator, their leading order calculation in
Section \ref{BDs3}, and their use to derive the Navier-Stokes
approximations to the heat and momentum fluxes. Of particular
interest are the long wavelength forms of the hydrodynamic
eigenfunctions $\xi _{\beta }^{h}$ given in Eq.\ (\ref {BD3.8}),
which are not simply linear combinations of the conserved
densities $1,\mathbf{v},v^{2}$.  It is clear from (\ref{BD3.06})
that perturbations of the homogeneous state due to $\xi _{\beta
}^{h}( \mathbf{c})$ generate the hydrodynamic excitations, while
the response to these excitations is measured by the fields $
\widetilde{a }_{\beta}^{h}(s)$. These fields are averages of the
biorthogonal set of functions $\bar{\xi}_{\beta }$, which are
linear combinations of the conserved densities
$1,\mathbf{v},v^{2}$, as required by the macroscopic balance
equations. As anticipated in the Introduction, this ``conjugate''
relationship between the sets $\left\{ \xi _{\beta }^{h}\right\} $
and $\left\{ \bar{ \xi}_{\beta }\right\} $ is a key difference
between normal and granular gases. For elastic collisions,  the
two sets of fields are the same. This has important consequences
for the derivation of transport properties as it has been
illustrated here. For example, the reduction leading to the
Navier-Stokes approximation in Eq.\ (\ref{BD4.202}), depends
critically on the fact that the functions $\xi _{\beta }^{h}$ are
eigenfunctions of $ \Lambda $. Any other choice for the initial
perturbation would lead to a different time dependence, mixing
microscopic and hydrodynamic excitations, and precluding direct
identification of transport properties.

The identification of the hydrodynamic modes discussed here is
important for the application of linear response methods
\cite{BDKTyH91} to granular systems. In order to illustrate it,
let us consider again the formal solution to the linear Boltzmann
equation given in (\ref{BD3.03}) and choose for the initial
condition the linearization of a \emph{local} HCS as given by Eq.\
(\ref{BD3.9c}),
\begin{eqnarray}
\delta \widetilde{\chi }(\mathbf{k},\mathbf{c},s)&=&e^{s\left(
\Lambda -i \mathbf{k}\cdot \mathbf{c}\right)} \sum_{\beta} <
\bar{\xi}_{\beta }^{h}|\delta \widetilde{\chi }(0)> \xi _{\beta
}^{h}({\bf c}) \notag \\
&=&e^{s\left( \Lambda -i\mathbf{k}\cdot \mathbf{c}\right)}
\sum_{\beta} a_{\beta }(\mathbf{k},s=0) \xi _{\beta }^{h}({\bf
c}). \label{5.1}
\end{eqnarray}
This initial state clearly represents a hydrodynamic perturbation.
The response $a_{\beta }(\mathbf{k},s)$ is then given by
\begin{equation}
a_{\beta }(\mathbf{k},s)=\sum_{\beta^{\prime}} R_{\beta
\beta^{\prime}}(\mathbf{k},s)a_{\beta^{\prime}}(\mathbf{k },s=0),
\quad R_{\beta \beta^{\prime}}(\mathbf{k},s)=< \bar{\xi}_{\beta
}^{h}|e^{s\left( \Lambda -i\mathbf{k}\cdot \mathbf{c}\right) }\xi
_{\beta^{\prime}}^{h}>. \label{5.2}
\end{equation}
The response functions $R_{\beta \beta^{\prime} }(\mathbf{k},s)$
are time correlation functions comprised of the pair of the
conjugate densities. It can be shown that application of standard
linear response methods to these response functions leads to the
correct Green-Kubo relations with the proper convergence
properties for the time integrals. Conversely, the same procedure
applied to correlation functions comprised of the $\left\{
\bar{\xi}_{\beta }\right\}$ alone does not lead directly to
well-posed Green-Kubo expressions. Let us also mention that linear
perturbations of the kind considered in Eq.\ (\ref{5.1}) have been
used \cite{BDBRyC99}in direct Monte Carlo simulations of the
Boltzmann equation \cite{BDBi94}, to verify the validity of the
linear hydrodynamic description and the explicit expressions of
the transport coefficients.

Although the analysis here has been limited to the low density
linearized Boltzmann equation, these considerations point the way
for a proper application of linear response methods in the more
general context of nonequilibrium statistical mechanics.

The research of J.W.D. reported here was supported in part by the
Department of Energy Grants DE-FG03-98DP00218 and DE-FG02ER54677.
The research of J.J.B. and M.J.R. was partially supported by the
Ministerio de Ciencia y Tecnolog\'{\i}a (Spain) through Grant No.
BFM2002-00307.

\bibliography{cecam02BIB}

\end{document}